\documentclass[twocolumn,aps,prb,showpacs]{revtex4}

\usepackage{epsf}
\usepackage{bm}

\begin{document}

\title {
         Dopant induced stabilization of Silicon cluster\\
         at finite temperature.
       }
\author {
         Shahab Zorriasatein, Kavita Joshi, and D. G. Kanhere
        }

\affiliation {
         Department of Physics,  and
         Center for Modeling and Simulation, 
         University of Pune, 
         Ganeshkhind, 
         Pune--411 007, 
         India 
             }

\date{\today}

\begin{abstract}
With the advances in miniaturization, understanding and controlling 
properties of significant technological systems like silicon in nano regime
 assumes considerable importance.
The small silicon clusters in the size range of 15-20 atoms are known to
 fragment upon heating. 
In the present work we demonstrate that it is possible to stabilize such
 clusters by introducing appropriate dopant (in this case Ti).
Specifically, by using the first principle density functional simulations we
 show  that Ti doped Si$_{16}$, having the Frank-Kasper geometry, remains
 stable till  2200~K and fragments only above 2600~K. The observed melting
 transition is a two step process. The first step is initiated by the surface
 melting around 600~K.
 The second step is the destruction of the cage which occurs around 2250~K
 giving rise to a peak in the heat capacity curve.
\end{abstract}

\maketitle

\section{Introduction}

 With the advent of nano science and  technology,  investigating physical
 properties of clusters has assumed considerable importance.~\cite{Reinhard}
  As the size of the system becomes  smaller and smaller its physical and chemical
  properties are expected to change,  sometimes significantly.
 In fact,  this is  what drives the quest for nano materials research.
 However, it may happen that  this size effect could drive a property towards
 an undesirable direction. The finite temperature properties of small silicon
 clusters is one such  example.
 In our recent finite  temperature studies~\cite{Sailaja} it was
 found that the  small clusters of silicon (Si$_{n}$, $n$=15 and 20) 
 become unstable and fragment when heated up
 to about 1500~K.
 Typically,  our extensive density functional simulations showed
 that Si$_{15}$ fragments around 1800~K in to Si$_{9}$ and Si$_{6}$, although we also  observe 
 Si$_{10}$+Si$_{5}$ and Si$_{8}$+Si$_{7}$ for a short time span.
 It exists in a  liquid like phase over a short temperature range
 (900 to 1400~K) before fragmenting. On the other hand,  Si$_{20}$  
 transforms from the ground state to the first  isomer (with two distinct Si$_{10}$
 units) around 1000~K and eventually fragments in to two Si$_{10}$ units around 1200~K. 
 Thus, Si$_{20}$ does not show any solid-like to liquid-like transition prior to
 fragmentation.
 The result  although not surprising,  can  have technological  implications,  
 silicon being the key  ingredient in most of the  semiconductor devices.
 The present work addresses the issue of remedying this  difficulty by using
 appropriate dopant.
 That the impurities induce significant changes in the structure as well as
 electronic properties is well known in the solid state community. In the 
 context of clusters several studies have been reported. For example,   single
 atom of tin or aluminum  in lithium  clusters are known to change the nature
 of bonding among the host atoms.~\cite{KJCP,Landman}
 In a recent work Mottet {\it et. al.}~\cite{Mottet} demonstrated that a single
 impurity (Ni or Cu) in Ag$_{55}$ cluster  changes structural and thermal
 properties dramatically.
 Interestingly, this effect was found to be  significant only on an icosahedral
 host. 

Silicon is the most important semiconducting materials in the
 microelectronics industry. In medium-sized regime, the structure and
 properties  of materials  often  differ dramatically from those of the bulk.
Over the past several years there have been extensive theoretical and
 experimental work describing the ground state geometries of medium sized
 silicon clusters.\cite{Ho,Rata,Lei,Jackson-2004,Si-PRL-Dynamics,PSM}
These clusters are built upon stable tricapped trigonal prism (TTP) units
which plays a crucial role in the finite temperature behavior of the cluster.\cite {PSM} The
 high stability of TTP units results in to fragmentation of these clusters
 instead of solid-like to liquid-like transition .\cite{Sailaja} The ground 
state  geometries of these clusters are well established in the literature.
 The ground  state geometry of Si$_{16}$ can be described as two fused
 pentagonal prisms.\cite{Lei}

Experiments on doped silicon clusters Si$_{n}$M ( M=Ti,  Hf, Cr, Mo, and
 W)  have shown large abundances for $n= 15, 16$  and low intensities
 for other sizes.\cite{EAS,ex-dop2,ex-dop3} Specifically the abundances
 drop drastically beyond  $n= 16$, giving support to the predictions of the
 exceptional stability  of $n= 16$ clusters.
 Vijay Kumar and co-workers~\cite{VJ2} showed that for $n= 8-12$
  basket-like open structures to be the most  favorable, while for $n= 13-16$,
 the metal atom is completely surrounded by silicon atoms. Recently,
 measurements of  electron  affinities of Ti-doped silicon clusters~\cite{EAS}
 have been  found to have  a minimum at $n= 16$, suggesting the closed electronic 
shell  nature and  strong stability of Si$_{16}$Ti.
 The electronic structure of silicon clusters in the size range noted
 above has been studied by {\it ab initio} methods.~\cite{Ho, Rata,Lei} It
 may be  noted that none of the silicon clusters in this size range form stable
 caged structures.
 In a quest to obtain stable  silicon cage structures, Vijay Kumar and
 co-workers~\cite{VJ1,VJ2,VJ3,VJ4} found that  it is possible  to stabilize a
 caged structure  for cluster using a class of dopants.
 Specifically, they showed that a single impurity of transition metal atoms like Ti, 
Zr, Hf enhances the binding energy of Si$_{16}$ and changes the structure to a
 caged one,  similar to carbon cages. The gain in binding energy is substantial
 and is  of the order of few eV and  HOMO-LUMO gap is more than 1 eV which
make these  structures stable. Among the systems studied,  Si$_{16}$Ti has
 the largest HOMO-LUMO gap (2.35~ eV) as compared to other dopants of
 Si$_{16}$ like Zr and Hf.~\cite{VJ1}
The ground state geometry for Si$_{16}$Ti has been found to be a Frank Kasper polyhedron,
  which is also shown in Fig.\ \ref{fig1}-a.  Therefore Si$_{16}$Ti is a likely
 candidate to  show a stable behavior at finite temperatures.
 As we shall  demonstrate, by adding a dopant like Ti, we could avoid the 
fragmentation observed in small Si clusters. The cluster shows a surface melting of Si atoms. 
However, the motion of Si atoms is restricted to a small shell around Ti until 2200~K.
The cluster is on the verge of fragmentation around 2600~K, almost 1000~K higher than that in pure Si clusters.
\section{Computational Details}

We have carried out isokinetic Born--Oppenheimer Molecular Dynamic (BOMD) simulations using  ultrasoft
 pseudopotentials within the Generalized Gradient  Approximation
 (GGA), as implemented in the {\rm VASP} package.\cite{GGA} For computing
 heat capacities, the BOMD calculations were carried out for 17 different 
 temperatures in  the range of  100~K to 3000~K, each with the  duration of
 150~ps or more, which  result in to a  total  simulation time of 2.6~ns. In
 order  to get converged  heat capacity curve  especially in the region of 
 coexistence,  more  temperatures were  required with longer simulation times.
 We have discarded  the first 30~ps of  each temperature for thermalization.
 To analyze the thermodynamic properties,  we first calculate the ionic 
 specific heat by using the Multiple Histogram (MH) technique.~\cite{Mh1, Mh2}
 We extract the  classical ionic density of states ($\Omega (E)$) of 
 the system, or equivalently  the classical ionic entropy, $S(E)=k_{B}\ln \Omega (E)$ following the MH technique. 
 With $S(E)$ in hand, one can evaluate thermodynamic averages in a variety of
ensembles.  We focus in this work on the ionic specific heat and the
caloric curve.  In the canonical ensemble, the specific heat is defined as
usual by $C(T)=\partial U(T)/\partial T$, where $U(T)=\int E\,p(E,T)\,dE$
is the average total energy, and where the probability of observing an
energy $E$ at a temperature $T$ is given by the Gibbs distribution
$p(E,T)=\Omega (E)\exp (-E/k_{B}T)/Z(T)$, with $Z(T) $ the normalizing
canonical partition function. 
 We normalize the calculated canonical  specific heat by the zero-temperature
 classical limit of the rotational  plus vibrational specific heat,
 i.e., $C_{0}=(3N-9/2)k_{B}$.

 We have calculated the root-mean-square bond length fluctuations
 ($\delta_{\rm rms}$) for Si-Si  and Si-Ti bonds separately to compare the
difference between them. The $\delta_{\rm rms}$ is  defined as
\begin{equation}
\delta _{{\rm rms}}=\frac{1}{N}\sum_{i>j}\frac{(\langle
r_{ij}^{2}\rangle _{t}-\langle r_{ij}\rangle _{t}^{2})^{1/2}}{\langle
r_{ij}\rangle _{t}},   \label{eqn:delta}
\end{equation}
where $N$ is the number of bonds in the system(N=120 for Si-Si,
and N=16 for Si-Ti),  $r_{ij}$ is the distance
between atoms $i$ and $j$,  and $\langle \ldots \rangle _{t}$ denotes a
time average over the entire trajectory. 
Mean Square Displacement (MSD)  is another traditional parameter
used for determining phase transition and is defined as,
\begin{equation}
\langle {\bf r}^{2}(t)\rangle =\frac{1}{NM}
\sum_{m=1}^{M}
\sum_{I=1}^{N}
\left[ {\bf R}_{I}(t_{0m}+t)-{\bf R}_{I}(t_{0m})\right]^{2}
\label{eqn:msq}
\end{equation}
here $N$ is the number of atoms in the system(N=16 for Si and N=1 for Ti)
  and R is the position
of the  $I$$^{th}$ atom.
Here we average over M different time origins t$_{0m}$ spanning
over the entire trajectory. The interval between the consecutive
t$_{0m}$ for the average was taken to be about 0.3~ps. The MSD of a cluster
 indicate the displacement of atoms in the cluster as a function of time.

 We have also calculated radial distribution function (g(r)) and deformation
 parameter ($\varepsilon_{def}$). 
 g(r)  is defined as  the average number of atoms within the region $r$ and
 $r+dr$.
The shape deformation parameter ($\varepsilon_{def}$) is defined as,
\begin{equation}
\varepsilon_{def} = \frac{2Q_{x}}{Q_y+Q_z}, 
\label{eqn:epspro}
\end{equation}
where
$Q_x \geq Q_y \geq Q_z$ are the eigenvalues,  in descending
order,  of the quadrupole tensor
\begin {equation}
        Q_{ij} =  {\sum_{I}  R_{Ii}\, R_{Ij} }.
\end {equation}
Here $i$ and $j$ run from 1 to 3,  $I$ runs over the number of ions,
and $R_{Ii}$ is the i$^{th}$ coordinate of ion $I$ relative to the
COM of the cluster. A spherical system($Q_{x}=Q_{y}=Q_{z}$) has 
$\varepsilon_{def}$=1 and larger values of $\varepsilon_{def}$ indicates deviation of the 
shape of the cluster from sphericity.
We use the electron localization function (ELF)~\cite{becke} to investigate the
 nature of bonding. For a single determinantal wave function built from
 Kohn-Sham orbitals $\psi _{i}$,  the  ELF is defined as
\begin{equation}
\chi _{{\rm ELF}}=[1+{(D/D}_{h}{)}^{2}]^{-1}, 
\end{equation}
where
\begin{eqnarray} D_{h}&=&(3/10){(3{\pi
}^{2})}^{5/3}{\rho }^{5/3},  \\ D&=&(1/2)\sum_{i}{\
|{\bm{\nabla} \psi _{i}}|}^{2}-(1/8){|{\bm{\nabla}
\rho }|}^{2}/\rho,  \end{eqnarray} with $\rho \equiv \rho
({\bf r})$ is the valence-electron density.
The ELF is defined in such a way that its value is unity for completely localized
systems and 0.5 for homogeneous electron gas.

\section {Results and Discussion}
\begin{figure}
  \epsfxsize=0.40\textwidth
  \centerline{\epsfbox{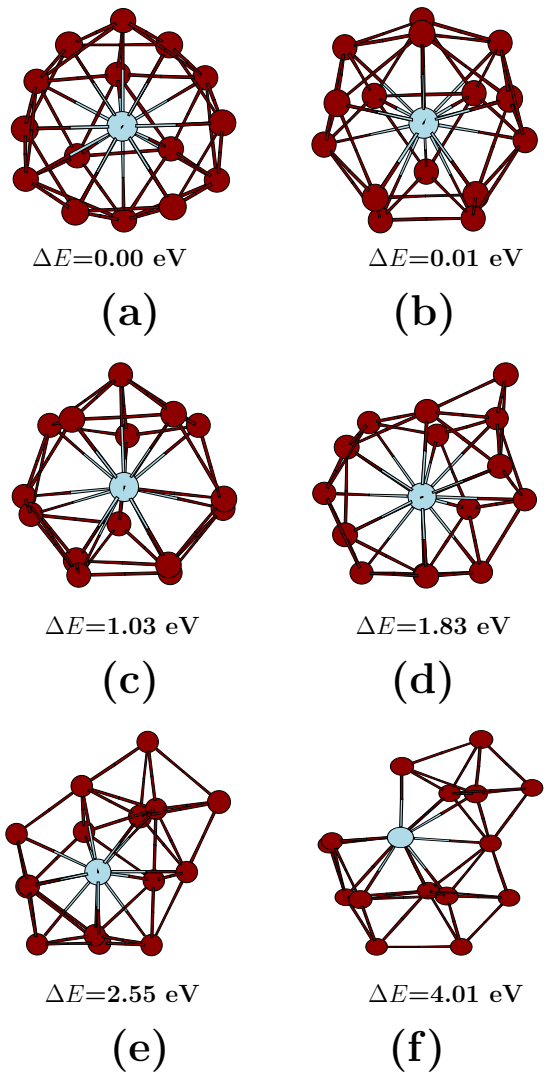}}
  \caption{\label{fig1}
  The ground state geometry and few interesting isomers of Si$_{16}$Ti
  }
\end{figure}
We begin our discussion by noting the ground state (GS) and some low lying structures 
of Si$_{16}$Ti which are shown in Fig.\ \ref{fig1}. 
As has been already mentioned the ground state of Si$_{16}$Ti is Frank-Kasper polyhedron.
The central Ti atom is surrounded by 16 Si atoms within two closely spaced 
shells one with 12 atoms, all equidistant from central Ti atom and another
 shell consisting of 4 Si atoms forming a tetrahedron with Ti at the center.
The first isomer (Fig.\ \ref{fig1}-b)  is energetically very close to the ground state where all  
Si atoms are nearly equidistant from central Ti atom ($\Delta$ E = .009 eV).
 Isomers with either distorted cage of Si atoms or without Si cage are quiet
high in energy compared to the lowest two structures. Isomer shown in
 Fig.\ \ref{fig1}-f occurs at very high temperature (around 2600~K) and indicates  
possible path for fragmentation.
At this point its interesting to compare pure silicon clusters with doped clusters.
Si clusters have very interesting growth pattern. The ground state 
geometries of Si$_{n}$ till $n=22$ are prolate and 
for larger sizes the GS transforms into spherical structures. It has been also
 observed that a TTP unit is a part of prolate
 ground states. Presence of highly stable TTP unit also bypass the liquid-like
 state and leads to fragmentation of pure Si$_{n}$ clusters around 1200~K to
 1800~K in the size range of 15-20.
Further, the ground state of Si$_{16}$ is a prolate structure and a cage like
 isomer is found to be  difficult to stabilize.

\begin{figure}
  \epsfxsize=0.4\textwidth
  \centerline{\epsfbox{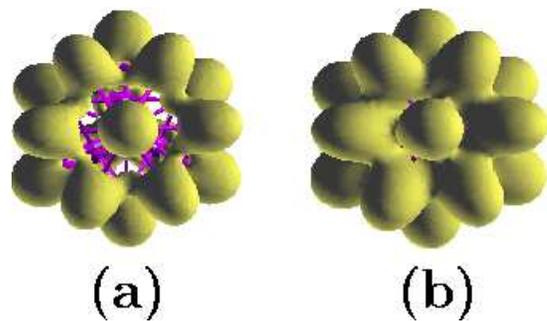}}
  \caption{\label{fig2}
The  isosurfaces of ELF at $\chi_{\rm ELF}$=.70 and $\chi_{\rm ELF}$=0.55 are
 shown in (a) and (b) respectively. 
 }
\end{figure}
The nature of bonding in Si$_{16}$Ti can be discussed by examining the iso-surfaces of
total charge density and ELF.
In Fig.\ \ref{fig2} we show the iso-surface of ELF for value of 0.7 
(Fig.\ \ref{fig2}-a) and 0.55 (Fig.\ \ref{fig2}-b).
 At higher values of $\chi_{\rm ELF}$ the figure 
clearly shows localization of the charge on hexagonal rings depicting the
covalent nature of bonding. These atoms are connected to each other 
by the shortest bonds (2.37-2.43~\AA) and are connected to the remaining four Si
atoms  at very low value of $\chi_{\rm ELF}$ $\approx$ 0.60
 ( bondlengths around 2.65~\AA). The bonding between silicon atoms belonging to two different shells 
 is metallic like in the sense it is due to
 delocalized charge distribution.

\begin{figure}
  \epsfxsize=0.5\textwidth
  \centerline{\epsfbox{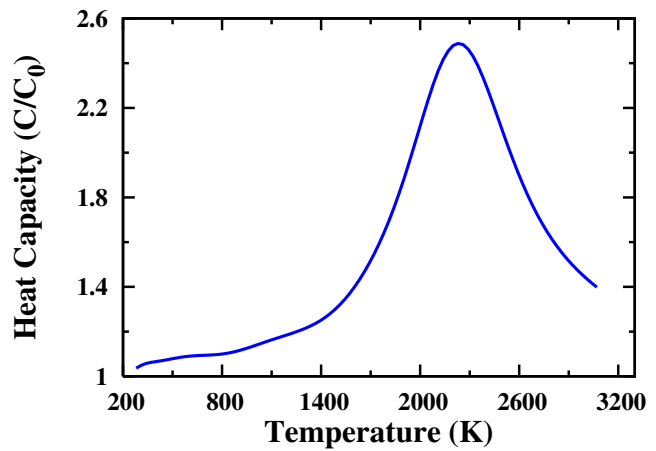}}
  \caption{\label{fig3}
  The heat capacity for Si$_{16}$Ti computed over last 120 ps. The peak is at 2250~K.
  }
\end{figure}
\begin{figure}
  \epsfxsize=0.45\textwidth
  \centerline{\epsfbox{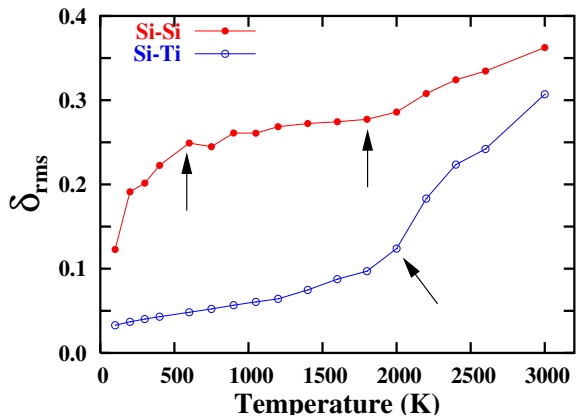}}
  \caption{\label{fig4}
  The $\delta_{\rm rms}$ for Si-Si bonds (online red color) and Si-Ti bonds
(online blue color) are shown. 
  The region between 600~K to 1800~K (shown by arrows for Si-Si bonds) 
 corresponds to `restricted' liquid-like behavior of Si cage.
  The rise in $\delta_{\rm rms}$ observed after 2000~K (shown by arrow for Si-Ti
 bonds) is associated with the breaking of Si cage 
  and diffusion of Ti through out the cluster.
  }
\end{figure}
\begin{figure}
  \epsfxsize=0.45\textwidth
  \centerline{\epsfbox{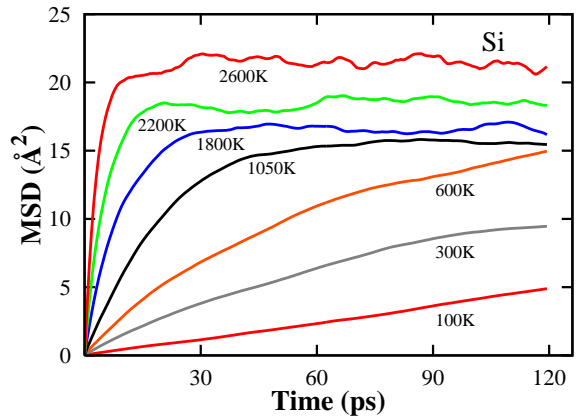}}
  \caption{\label{fig5} The MSD calculated for Si atoms for few relevant temperatures.
  }
\end{figure}
\begin{figure}
  \epsfxsize=0.45\textwidth
  \centerline{\epsfbox{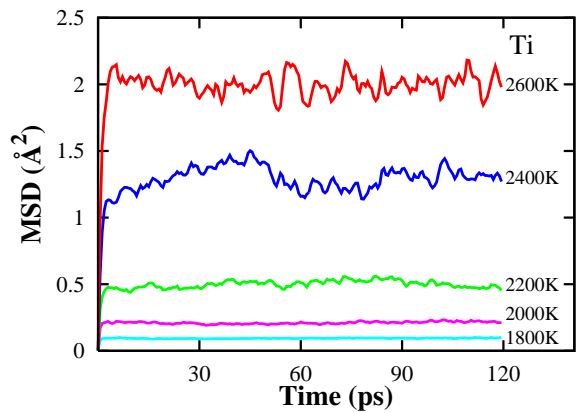}}
  \caption{\label{fig6} The MSD calculated for Ti atom for five different temperature
  }
\end{figure}

Next we show the ionic heat capacity of Si$_{16}$Ti in Fig.\ \ref{fig3}.
The heat capacity displays a broad melting peak at 2250~K characteristic of a finite
size system. A more careful observation of ionic motions and traditional
 parameters like $\delta_{\rm rms}$ reveals that the main peak in the
 heat capacity is related to complete breakdown of Si cage and 
escape of Ti atom from the cage. The $\delta_{\rm rms}$ for Si-Si and Si-Ti 
bonds,  shown in Fig.\ \ref{fig4}, throws light on the finite temperature
 behavior of this cluster. The value of $\delta_{\rm rms}$ for Si-Si bonds 
is considerably high at much lower temperatures (around 200~K) whereas the value 
of $\delta_{\rm rms}$ for Si-Ti bonds is quite low till 
1800~K. The higher value of $\delta_{\rm rms}$ for Si atoms indicates that Si atoms are mobile at low temperatures.
More detailed analysis of ionic motion reveals that the cluster undergoes isomerization around as 
low temperature as 100~K. The cluster
transforms from the GS to the first isomer and back to the GS which in turn results in to diffusion of at least 4  atoms. 
The MSD also support this observation. In Fig.\ \ref{fig5} and Fig.\ \ref{fig6} 
the MSD computed for Si and Ti atoms at some relevant temperatures 
are shown respectively. As can be noted from Fig.\ \ref{fig5} the MSD for Si atoms saturate around 600~K indicating that 
the atoms are diffusing on the spherical shell around Ti atom and the cluster is partially melted. 
On the contrary, as can be seen from Fig.\ \ref{fig6} the Ti atom
shows very small displacement till 2000~K. 
Although the MSD shows liquid-like behavior around much lower temperature (600~K) 
the Si atoms are confined to a small shell around Ti (preserving the
 spherical shape) until quite high temperatures. Its only around 
2000~K or so that the Si cage breaks down giving rise to a peak in the heat
 capacity around 2200~K. That the motion of Si atoms is restricted to
a shell is clear from the examination of radial distribution function.
\begin{figure}
  \epsfxsize=0.45\textwidth
  \centerline{\epsfbox{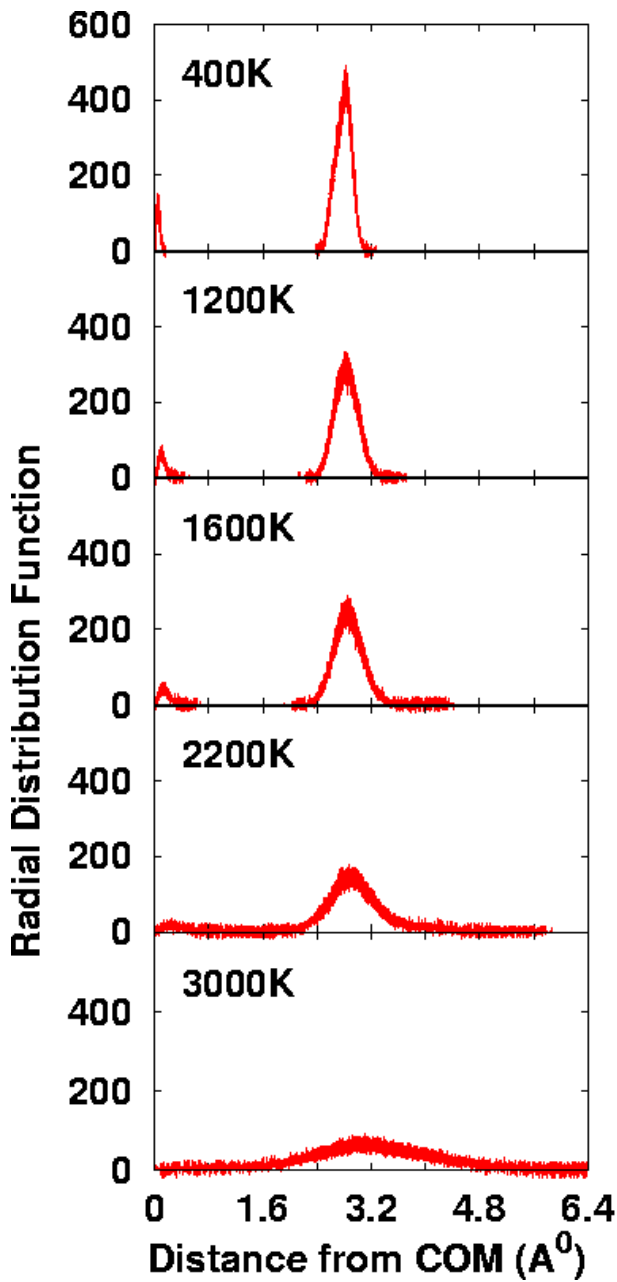}}
  \caption{\label{fig7} The radial distribution function calculated for five different temperatures.
  }
\end{figure}
In Fig.\ \ref{fig7} we show the radial
 distribution function calculated from the center of mass (Ti site) for
some representative temperatures. The peak near origin is due to the Ti atom whereas
 a single peak around 2.8~\AA\ is due to Si cage. The width of this peak does not
 change significantly till 1600~K. A continuous distribution emerges after 2200~K
 indicating that the Si cage is destroyed and Ti has escaped from the cage.  
This change is accompanied by the   shape change of the cluster.

\begin{figure}[t]
\vskip 0.8 cm
  \epsfxsize=0.45\textwidth
  \centerline{\epsfbox{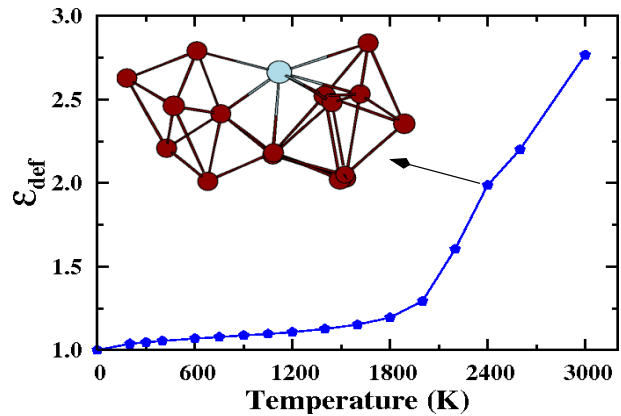}}
  \caption{\label{fig8} The deformation parameter is plotted as a function of temperature. 
   Note the sudden rise in the deformation parameter after 2000~K. A typical
 configuration shown in the Fig occurs around 2400~K, and indicates possible
path for fragmentation.
  }
\end{figure}
In Fig.\ \ref{fig8} we show shape deformation parameter as a function of temperature. 
It  can be seen that till 2000~K value of $\varepsilon_{def}$ is nearly 
1 indicating a spherical shape with restricted motion of Si atoms on the surface. 
 The $\varepsilon_{def}$ increases  significantly above 2000~K  
indicating the breakdown of Si cage. Our simulation shows that a fragmentation
 channel opens up around 2600~K possibly because of emergence of central Ti
 atom on the surface. We have observed that Si$_{16}$Ti around this temperature
  fragments in to two parts, the most probable fragments are   Si$_{12}$Ti+Si$_{4}$  and Si$_{13}$Ti+Si$_{3}$.   
A typical configuration leading to fragmentation of Si$_{16}$Ti 
is also shown in the Fig.\ \ref{fig8}.

\section{Summary and conclusion\label{sec:concl}}

To summarize, our DFT simulations clearly demonstrate that its possible 
to avoid fragmentation of small Si clusters by doping them with appropriate
impurity, in this case Ti. Contrary to the pure Si clusters, the doped cluster
undergoes a solid-like to liquid-like transition, and remain  stable at least 
up to 2600~K.
However, the `melting' occur in two steps, the first one is a  low 
temperature process (around 600~K), where Si atoms diffuse on the shell,
whereas the peak in the heat capacity (2250~K) is associated with the breaking 
of Si cage and escape of Ti atom from the cage. 
The beginning of the fragmentation is observed around 2600~K 
which is almost 1000~K higher than that observed in pure Si clusters in this size
range.

\section  {Acknowledgment}
KJ and DGK thank Indo French Center for Promotion of Advanced Research
(IFCPAR) for partial financial support(project No; 3104-2). The authors also thanks M--S Lee
for many useful discussions.


\begin{thebibliography}{}
\bibitem{Reinhard}
P. G. Reinhard, E . Suraud,
{\it Introduction to Cluster Dynamics} (Wiley-VCH, 2004).
\bibitem{Sailaja}
S. Krishnamurty, K. Joshi,  D. G. Kanhere, and S. A. Blundell,
  Phys. Rev. B. {\bf 73}, 045419 (2006).
\bibitem{KJCP}
K. Joshi and  D. G. Kanhere,
  J. Chem. Phys. {\bf 119}, 12301 (2003).
M.--S. Lee, D. G. Kanhere, and K. Joshi,
 Phys. Rev. A. {\bf 72}, 015201 (2005).
\bibitem{Landman}
  H.--P. Cheng, R. N. Barnett, and  U. Landman,
 Phys. Rev. B. {\bf 48}, 1820 (1993).
\bibitem{Mottet}
C. Mottet, G. Rossi, F. Baletto, and R. Ferrando,
 Phys. Rev. Lett. {\bf 95}, 035501 (2005).
\bibitem{Ho}
K.-M. Ho, A.A. Shvartsburg, B. Pan, Z.Y. Lu, C.Z
                      Wang, J.G. Wacker, J.L. Fye, and M.F. Jarrold,
                      Nature, {\bf 392}, 582 (1998)
\bibitem{Rata}
 I. Rata, A. A. Shvartsburg, M. Horoi, T.Frauenheim,
                    Siu K. W. Michael, and K. A. Jackson
                    Phys. Rev. Lett. {\bf 85}, 546 (2000).
\bibitem{Lei}
X. L. Zhu, X. C. Zeng, Y. A. Lei, and B. Pan
                    J. Chem. Phys. {\bf 120}, 8985 (2004).
\bibitem{Jackson-2004}
 K.A. Jackson, M. Horoi, I. Chaudhuri,
                T. Frauenheim, and A. A. Shvartsburg
                Phys. Rev. Lett. {\bf 93}, 013401, (2003).
\bibitem{Si-PRL-Dynamics}
L. Mitas, J.C. Grossman, I. Stich, and J. Tobik 
                          Phys. Rev. Lett. {\bf 84}, 1479 (2000)
\bibitem{PSM}
J. Muller, B. Liu, A.A. Shavartsburg, S. Ogut,
    J. R. Chelikowsky, Siu K. W. Michael, K. M. Ho, and G. Gantefor
    Phys. Rev. Lett.  {\bf 85}, 1666 (2000)
\bibitem{EAS}
 M. Ohara, K. Koyasu, A. Nakajima, and K. Kaya,
 Chem. Phys. Lett {\bf 371}, 490 (2003).
\bibitem{ex-dop2}
P. Sen and  L. Mitas,
Phys. Rev.B {\bf 68}, 155404 (2003).
\bibitem{ex-dop3}
V. Kumar, T. M. Briere, and  Y. Kawazoe,
Phys. Rev.B {\bf 68}, 155412 (2003).
\bibitem{VJ1}
V. Kumar and  Y. Kawazoe,
Phys. Rev. Lett. {\bf 87}, 045503 (2001).
 \bibitem{VJ2}
H. Kawamura, V. Kumar, and  Y. Kawazoe, 
Phys. Rev.B {\bf 71}, 075423 (2005). 
\bibitem{VJ3}
V. Kumar and  Y. Kawazoe,
 Phys. Rev.B {\bf 65}, 073404 (2002). 
\bibitem{VJ4}
V. Kumar, A. K.Singh, and  Y. Kawazone,
Nano Lett. {\bf 4}, 677 (2004).
\bibitem{GGA}
Vienna {\em ab initio } simulation package,
                     Technische Universit\"at Wien (1999);
                     G. Kresse and J. Furthm\"uller,
                     Phys. Rev. B {\bf 54}, 11169 (1996).
\bibitem{Mh1}
A. M. Ferrenberg, R. H. Swendsen,
 Phys. Rev. Lett. {\bf 61}, 2635 (1988).
\bibitem{Mh2}
P. Labastie and  R. L. Whetten 
{\it  Phys. Rev. Lett.} {\bf 65}, 1567 (1990).  
\bibitem{becke}    
 A. D. Becke and  K. E. Edgecombe,
 J. Chem. Phys. {\bf 92}, 5397 (1990).

\end{thebibliography}
\end{document}